\documentclass[aps,prd,twocolumn,superscriptaddress,showpacs,floatfix]{revtex4}

\usepackage{epsfig}
\usepackage{psfrag}
\usepackage{graphicx}
\usepackage{dcolumn}
\usepackage{bm}
\usepackage{epsfig}
\usepackage{lineno}
\usepackage{multirow}
\usepackage{slashed}
\usepackage{rotating}
\usepackage{subfigure}
\usepackage{caption}
\usepackage{verbatim} 
\usepackage{mathrsfs} 

\newcommand{\pt}{\mbox{$p_{T}$}~}
\newcommand{\pte}{$p_{T}$} 
\newcommand{\mjl}{\mbox{$D_{\text{MJL}}$}~}
\newcommand{\mjle}{\mbox{$D_{\text{MJL}}$}}

\newcommand{\dphi}{$\Delta\varphi_{Z,\text{jet}}$~}
\newcommand{\dphie}{$\Delta\varphi_{Z,\text{jet}}$}

\newcommand{\dzero} {\mbox{D0}~}
\newcommand{\DO}{\mbox{D0}~}

\newcommand{\GeV} {\ensuremath{\mathrm{Ge\kern -0.1em V}}~}
\newcommand{\GeVe} {\ensuremath{\mathrm{Ge\kern -0.1em V}}}
\newcommand{\TeV} {\ensuremath{\mathrm{Te\kern -0.1em V}}}

\newcommand{\ppbar}{\mbox{$p\overline{p}$}~}

\newcommand{\alpgen}{{\sc alpgen}~}
\newcommand{\alpgene}{{\sc alpgen}}

\newcommand{\mcfm}{{\sc mcfm}~}
\newcommand{\mcfme}{{\sc mcfm}}

\newcommand{\sherpa}{{\sc sherpa}~}
\newcommand{\sherpae}{{\sc sherpa}}

\newcommand{\ptj}{\mbox{$p_{T}^{\text{jet}}$}}
\newcommand{\ptz}{\mbox{$p_{T}^{Z}$}}
\newcommand{\etaj}{\mbox{$\eta^{\text{jet}}$}}

\newcommand{\ratio}{\mbox{$\sigma(Z+b~\text{jet})/\sigma(Z+\text{jet})$}}

\def\gsim{\mathrel{\rlap{\raise.4ex\hbox{$>$}} {\lower.6ex\hbox{$\sim$}}}}
\def\lsim{\mathrel{\rlap{\raise.4ex\hbox{$<$}} {\lower.6ex\hbox{$\sim$}}}}

\begin{document}

\hspace{5.2in} \mbox{FERMILAB-PUB-13-011-E}

\title{\boldmath Measurement of the ratio of differential cross sections\\ 
$\sigma (p\bar{p} \rightarrow Z+b~\text{jet}) / \sigma (p\bar{p} \rightarrow Z+ \text{jet})$ 
in $p\bar{p}$ collisions at $\sqrt s=1.96~\TeV$}

\affiliation{LAFEX, Centro Brasileiro de Pesquisas F\'{i}sicas, Rio de Janeiro, Brazil}
\affiliation{Universidade do Estado do Rio de Janeiro, Rio de Janeiro, Brazil}
\affiliation{Universidade Federal do ABC, Santo Andr\'e, Brazil}
\affiliation{University of Science and Technology of China, Hefei, People's Republic of China}
\affiliation{Universidad de los Andes, Bogot\'a, Colombia}
\affiliation{Charles University, Faculty of Mathematics and Physics, Center for Particle Physics, Prague, Czech Republic}
\affiliation{Czech Technical University in Prague, Prague, Czech Republic}
\affiliation{Center for Particle Physics, Institute of Physics, Academy of Sciences of the Czech Republic, Prague, Czech Republic}
\affiliation{Universidad San Francisco de Quito, Quito, Ecuador}
\affiliation{LPC, Universit\'e Blaise Pascal, CNRS/IN2P3, Clermont, France}
\affiliation{LPSC, Universit\'e Joseph Fourier Grenoble 1, CNRS/IN2P3, Institut National Polytechnique de Grenoble, Grenoble, France}
\affiliation{CPPM, Aix-Marseille Universit\'e, CNRS/IN2P3, Marseille, France}
\affiliation{LAL, Universit\'e Paris-Sud, CNRS/IN2P3, Orsay, France}
\affiliation{LPNHE, Universit\'es Paris VI and VII, CNRS/IN2P3, Paris, France}
\affiliation{CEA, Irfu, SPP, Saclay, France}
\affiliation{IPHC, Universit\'e de Strasbourg, CNRS/IN2P3, Strasbourg, France}
\affiliation{IPNL, Universit\'e Lyon 1, CNRS/IN2P3, Villeurbanne, France and Universit\'e de Lyon, Lyon, France}
\affiliation{III. Physikalisches Institut A, RWTH Aachen University, Aachen, Germany}
\affiliation{Physikalisches Institut, Universit\"at Freiburg, Freiburg, Germany}
\affiliation{II. Physikalisches Institut, Georg-August-Universit\"at G\"ottingen, G\"ottingen, Germany}
\affiliation{Institut f\"ur Physik, Universit\"at Mainz, Mainz, Germany}
\affiliation{Ludwig-Maximilians-Universit\"at M\"unchen, M\"unchen, Germany}
\affiliation{Fachbereich Physik, Bergische Universit\"at Wuppertal, Wuppertal, Germany}
\affiliation{Panjab University, Chandigarh, India}
\affiliation{Delhi University, Delhi, India}
\affiliation{Tata Institute of Fundamental Research, Mumbai, India}
\affiliation{University College Dublin, Dublin, Ireland}
\affiliation{Korea Detector Laboratory, Korea University, Seoul, Korea}
\affiliation{CINVESTAV, Mexico City, Mexico}
\affiliation{Nikhef, Science Park, Amsterdam, the Netherlands}
\affiliation{Radboud University Nijmegen, Nijmegen, the Netherlands}
\affiliation{Joint Institute for Nuclear Research, Dubna, Russia}
\affiliation{Institute for Theoretical and Experimental Physics, Moscow, Russia}
\affiliation{Moscow State University, Moscow, Russia}
\affiliation{Institute for High Energy Physics, Protvino, Russia}
\affiliation{Petersburg Nuclear Physics Institute, St. Petersburg, Russia}
\affiliation{Instituci\'{o} Catalana de Recerca i Estudis Avan\c{c}ats (ICREA) and Institut de F\'{i}sica d'Altes Energies (IFAE), Barcelona, Spain}
\affiliation{Uppsala University, Uppsala, Sweden}
\affiliation{Lancaster University, Lancaster LA1 4YB, United Kingdom}
\affiliation{Imperial College London, London SW7 2AZ, United Kingdom}
\affiliation{The University of Manchester, Manchester M13 9PL, United Kingdom}
\affiliation{University of Arizona, Tucson, Arizona 85721, USA}
\affiliation{University of California Riverside, Riverside, California 92521, USA}
\affiliation{Florida State University, Tallahassee, Florida 32306, USA}
\affiliation{Fermi National Accelerator Laboratory, Batavia, Illinois 60510, USA}
\affiliation{University of Illinois at Chicago, Chicago, Illinois 60607, USA}
\affiliation{Northern Illinois University, DeKalb, Illinois 60115, USA}
\affiliation{Northwestern University, Evanston, Illinois 60208, USA}
\affiliation{Indiana University, Bloomington, Indiana 47405, USA}
\affiliation{Purdue University Calumet, Hammond, Indiana 46323, USA}
\affiliation{University of Notre Dame, Notre Dame, Indiana 46556, USA}
\affiliation{Iowa State University, Ames, Iowa 50011, USA}
\affiliation{University of Kansas, Lawrence, Kansas 66045, USA}
\affiliation{Kansas State University, Manhattan, Kansas 66506, USA}
\affiliation{Louisiana Tech University, Ruston, Louisiana 71272, USA}
\affiliation{Northeastern University, Boston, Massachusetts 02115, USA}
\affiliation{University of Michigan, Ann Arbor, Michigan 48109, USA}
\affiliation{Michigan State University, East Lansing, Michigan 48824, USA}
\affiliation{University of Mississippi, University, Mississippi 38677, USA}
\affiliation{University of Nebraska, Lincoln, Nebraska 68588, USA}
\affiliation{Rutgers University, Piscataway, New Jersey 08855, USA}
\affiliation{Princeton University, Princeton, New Jersey 08544, USA}
\affiliation{State University of New York, Buffalo, New York 14260, USA}
\affiliation{University of Rochester, Rochester, New York 14627, USA}
\affiliation{State University of New York, Stony Brook, New York 11794, USA}
\affiliation{Brookhaven National Laboratory, Upton, New York 11973, USA}
\affiliation{Langston University, Langston, Oklahoma 73050, USA}
\affiliation{University of Oklahoma, Norman, Oklahoma 73019, USA}
\affiliation{Oklahoma State University, Stillwater, Oklahoma 74078, USA}
\affiliation{Brown University, Providence, Rhode Island 02912, USA}
\affiliation{University of Texas, Arlington, Texas 76019, USA}
\affiliation{Southern Methodist University, Dallas, Texas 75275, USA}
\affiliation{Rice University, Houston, Texas 77005, USA}
\affiliation{University of Virginia, Charlottesville, Virginia 22904, USA}
\affiliation{University of Washington, Seattle, Washington 98195, USA}
\author{V.M.~Abazov} \affiliation{Joint Institute for Nuclear Research, Dubna, Russia}
\author{B.~Abbott} \affiliation{University of Oklahoma, Norman, Oklahoma 73019, USA}
\author{B.S.~Acharya} \affiliation{Tata Institute of Fundamental Research, Mumbai, India}
\author{M.~Adams} \affiliation{University of Illinois at Chicago, Chicago, Illinois 60607, USA}
\author{T.~Adams} \affiliation{Florida State University, Tallahassee, Florida 32306, USA}
\author{G.D.~Alexeev} \affiliation{Joint Institute for Nuclear Research, Dubna, Russia}
\author{G.~Alkhazov} \affiliation{Petersburg Nuclear Physics Institute, St. Petersburg, Russia}
\author{A.~Alton$^{a}$} \affiliation{University of Michigan, Ann Arbor, Michigan 48109, USA}
\author{A.~Askew} \affiliation{Florida State University, Tallahassee, Florida 32306, USA}
\author{S.~Atkins} \affiliation{Louisiana Tech University, Ruston, Louisiana 71272, USA}
% ...
%
%
\collaboration{The D0 Collaboration\footnote{with visitors from
$^{a}$Augustana College, Sioux Falls, SD, USA,
$^{b}$The University of Liverpool, Liverpool, UK,
$^{c}$UPIITA-IPN, Mexico City, Mexico,
$^{d}$DESY, Hamburg, Germany,
$^{e}$SLAC, Menlo Park, CA, USA,
$^{f}$University College London, London, UK,
$^{g}$Centro de Investigacion en Computacion - IPN, Mexico City, Mexico,
$^{h}$ECFM, Universidad Autonoma de Sinaloa, Culiac\'an, Mexico
and
$^{i}$Universidade Estadual Paulista, S\~ao Paulo, Brazil.
}} \noaffiliation
\vskip 0.25cm

\date{January 10, 2013}

\begin{abstract}
We measure the ratio of cross sections,
$\sigma(p\bar{p}\rightarrow Z+b~\text{jet})$/$\sigma(p\bar{p}\rightarrow Z+\text{jet})$,
for associated production of a $Z$ boson with at least one jet. 
The ratio is also measured as a function of the $Z$ boson transverse 
momentum, jet transverse momentum, jet pseudorapidity, and the 
azimuthal angle between the $Z$ boson with respect to the highest \pt 
$b$ tagged jet. These measurements use data collected by the \DO experiment in 
Run~II of Fermilab's Tevatron  \ppbar Collider at a center-of-mass energy 
of 1.96 TeV, and correspond to an integrated luminosity of 9.7 fb$^{-1}$.
The results are compared to predictions from next-to-leading 
order calculations and various Monte Carlo event generators. 
\end{abstract}

\maketitle

\newpage
Studies of $Z$ boson production in association 
with jets from $b$ quarks, or $b$ jets, provide important tests of the predictions of
perturbative quantum chromodynamics (pQCD)~\cite{Campbell}. 
A good theoretical description of this process is essential since it 
forms a major background for a variety of physics processes, including
the standard model (SM) Higgs boson production in association with
a $Z$ boson, $ZH(H\rightarrow b\bar{b})$~\cite{zhllbb}, and
searches for supersymmetric partners of the $b$ quark~\cite{sbottom}.
Furthermore, $Z+b~\text{jet}$ production can serve as a reference process
for a non-SM Higgs boson ($h$) produced in association with a $b$ quark.
Two different approaches are currently available to calculate $Z$
or $h$ boson production in association 
with a $b$ quark at next-to-leading order (NLO)~\cite{Campbell,Doreen}.
They yield consistent results within theoretical uncertainties~\cite{DawsonDoreen}.

The ratio of $Z+b~\text{jet}$ to $Z+\text{jet}$ production cross sections, 
for events with one or more jets, has been
previously measured by the CDF~\cite{CDFPaper,CDFPaperII} and 
\dzero~\cite{Zb_PRL, Zb_PRD} collaborations using a fraction of the Run~II data.
The ATLAS~\cite{atlas} and CMS~\cite{cms} collaborations have also 
examined $Z+b~\text{jet}$ production at $\sqrt{s} = 7$~TeV.
The results obtained by the experiments agree, within 
experimental uncertainties, with the theoretical predictions.

The current measurement is based on the complete Run~II
data sample collected by the D0 experiment~\cite{d0det} at
Fermilab's Tevatron \ppbar collider running with a center-of-mass
energy of $\sqrt{s}=1.96$~TeV, and corresponds to an integrated luminosity of 9.7 fb$^{-1}$.
The enlarged data 
sample enables the measurement of the cross section ratio,   
\ratio, to be performed differentially as a function of various 
kinematic variables. The $Z$ bosons are required to decay 
to pairs of leptons, $\mu\mu$ or $ee$, which pass at least
one of the single electron or muon triggers.
For our off-line event selection, the triggers have an
efficiency of approximately 100\% for $Z\rightarrow ee$ and more
than 78\% for $Z\rightarrow\mu\mu$ decays depending on the transverse momentum of the muon. 
The $Z+\text{jet}$ sample requires the presence of
at least one jet in the event, while the $Z+b~\text{jet}$ sample
requires at least one $b$-jet candidate, selected using a
$b$-tagging algorithm~\cite{bid}.
The measurement of the ratio of cross sections benefits from
nearly complete cancellation of several systematic uncertainties such as those associated with the
identification of leptons, jets, measurement of the luminosity, etc., and therefore allows for a more
precise comparison of data with various theoretical predictions.

This analysis relies on all components of the \dzero detector: 
tracking systems, liquid-argon sampling calorimeter, muon systems, 
and the ability to identify secondary vertices~\cite{d0det}. 
The silicon microstrip tracker (SMT) allows for precise reconstruction 
of the primary \ppbar interaction vertex~\cite{PV} 
and secondary vertices. It also enables  
an accurate determination of the impact parameter, defined as
a distance of closest approach of a track to the interaction vertex. The impact
parameter measurements of tracks, along with the secondary vertices, 
are important inputs to the $b$-tagging algorithm. 
A detailed description of the \dzero detector can be found elsewhere~\cite{d0det}.

An event is selected if it contains a \ppbar interaction vertex, built from 
at least three tracks, located within 60 cm of the center of 
the \DO detector along the beam axis.
The selected events must also contain a $Z$ boson candidate with a
dilepton invariant mass $70~<M_{\ell\ell}<110~\GeVe~(\ell = e,\mu)$.

Dielectron ($ee$) events are required to have two electrons of transverse 
momentum $\pt>15~\GeVe$ identified through electromagnetic (EM) showers in the calorimeter. 
The showers must have more than 90\% of their energy deposited in the EM
calorimeter, be isolated from other energy depositions, and have a transverse 
and longitudinal profile consistent with that expected for an electron. 
At least one electron must be identified in the central calorimeter (CC), within a pseudorapidity~\cite{def} region $|\eta|<1.1$,
and a second electron either in the CC or the endcap calorimeters, $1.5<|\eta|<2.5$. 
Electron candidates in the CC region are also required to match central tracks 
or have a pattern of hits consistent with the passage of an electron through the 
central tracker. There is no requirement on the charge of the selected electrons.

Dimuon ($\mu\mu$) events are required to have two oppositely charged 
muons detected in the muon spectrometer that are matched to
central tracks with \pt$>10~\GeVe$ and $|\eta|<2$. At least one muon 
is required to have \pt$>15~\GeVe$. These muons must pass a combined 
tracking and calorimeter isolation requirement.
Muons originating from cosmic rays are rejected by 
applying timing criteria using the hits in the scintillator layers and by 
limiting the measured displacement of the muon track with respect to the \ppbar interaction vertex.

A total of 1,249,911 $Z$ boson candidate events 
are retained in the combined $ee$ and $\mu\mu$ channels with the above
criteria. The $Z+\text{jet}$ sample is then selected by requiring at 
least one jet in the event with a corrected \ptj~$>20~\GeVe$ and $|\etaj|<2.5$.
Jets are reconstructed from energy 
deposits in the calorimeter using an iterative midpoint cone algorithm~\cite{RunIIcone} 
with a cone of radius
$\Delta R =  \sqrt{(\Delta\varphi)^{2}+(\Delta y)^{2}}= 0.5$ where
$\varphi$ is the azimuthal angle and $y$ is the rapidity.
Jet energy is corrected for detector response, the presence of noise, 
multiple $p\bar{p}$ interactions, and energy deposited outside 
of the jet cone used for reconstruction~\cite{JES}. 

To suppress background from top quark production, events are rejected if
the missing transverse energy is larger than $60~\GeVe$, reducing 
the $t\bar{t}$ contamination by a factor of two. These selection criteria 
retain an inclusive sample of 176,498 $Z+\text{jet}$ event candidates in 
the combined $ee$ and $\mu\mu$ channel.

Processes such as diboson ($WW$, $WZ$, $ZZ$) production can
contribute to the background when 
two leptons are reconstructed in the final state.
Inclusive diboson production is simulated with the
{\sc pythia}~\cite{pythia} Monte Carlo (MC) event generator.
The $Z+\text{jet}$, including heavy flavor jets, and $t\bar{t}$ events are modeled by
{\sc alpgen}~\cite{alpgen}, which generates hard sub-processes including higher 
order QCD tree level matrix elements, interfaced with
{\sc pythia} for parton showering and hadronization.
The {\sc CTEQ6L1}~\cite{cteq6} 
parton distribution functions (PDFs) are used in all simulations.
The cross sections of the simulated samples are then scaled to the corresponding 
higher order theoretical calculations. 
For the diboson and $Z+\text{jet}$ processes, 
including the $Z + b\bar{b}$ signal process and $Z + c\bar{c}$
production, next-to-leading order (NLO) cross section
predictions are taken from {\sc mcfm}~\cite{diboson}.
The $t\bar{t}$ cross section is determined from approximate next-to-NLO
calculations ~\cite{ttbar}. 
To improve the modeling of the \pt distribution of the $Z$ boson, simulated 
$Z+\text{jet}$ events are also reweighted to be consistent with the measured \pt 
spectrum of $Z$ bosons observed in data~\cite{zpt}. 

These generated samples are processed 
through a detailed detector simulation based on {\sc geant}~\cite{geant}. 
To model the effects of detector noise 
and pile-up events, collider data from random beam crossings are 
superimposed on simulated events.
These events are then reconstructed using the same algorithms as used for data.
Scale factors, determined from data using independent samples, are applied to 
account for differences in reconstruction efficiency between data and simulation. 
The energies of simulated jets are corrected, based on their flavor, to 
reproduce the resolution and energy scale observed in data~\cite{JES}.  
In the following, light-quark flavor ($u$, $d$, $s$) and gluon jets are referred to as ``light jets'' or ``LF''.

\begin{figure}
\centering
\includegraphics[width=0.5\columnwidth]{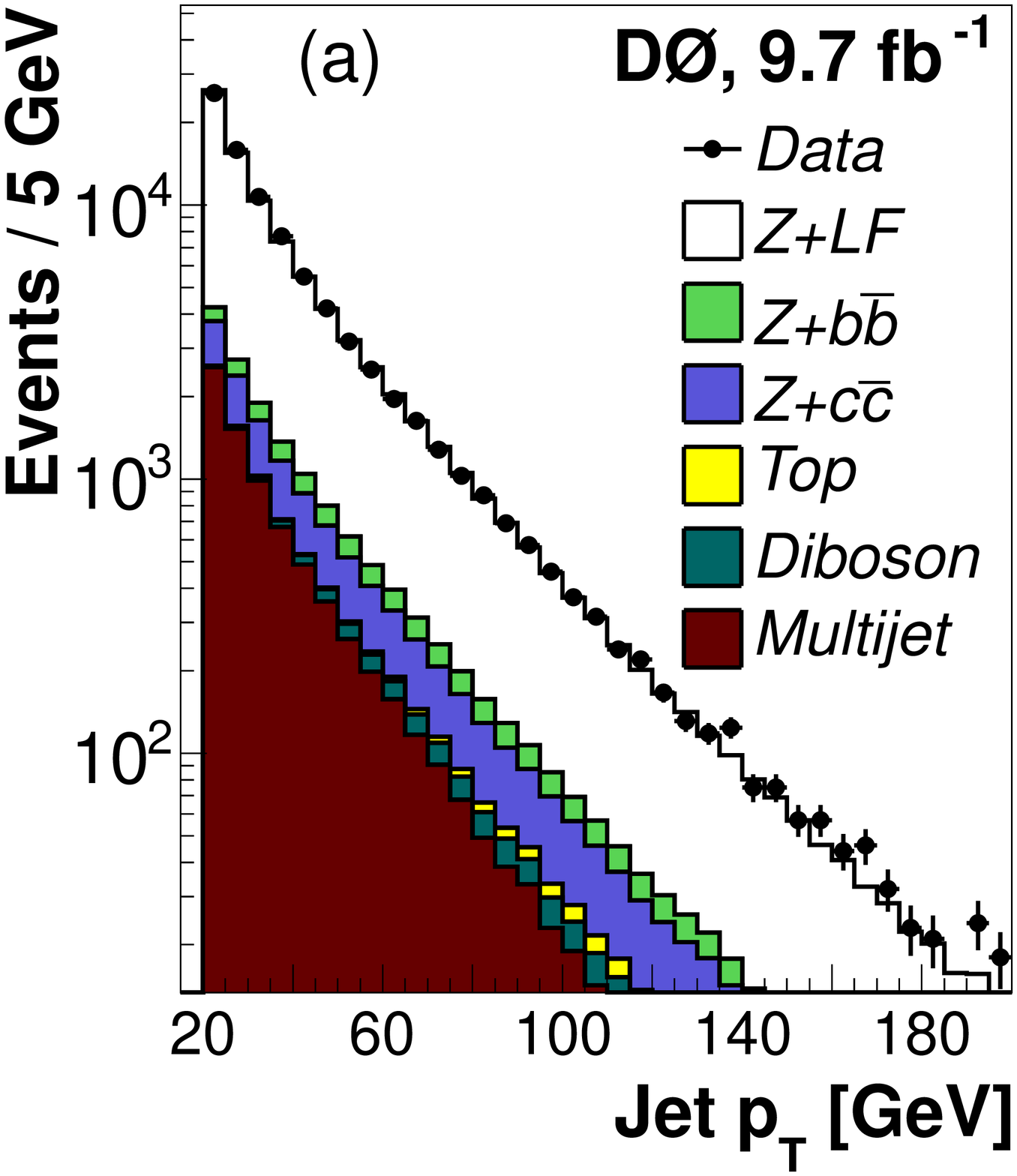}\includegraphics[width=0.5\columnwidth]{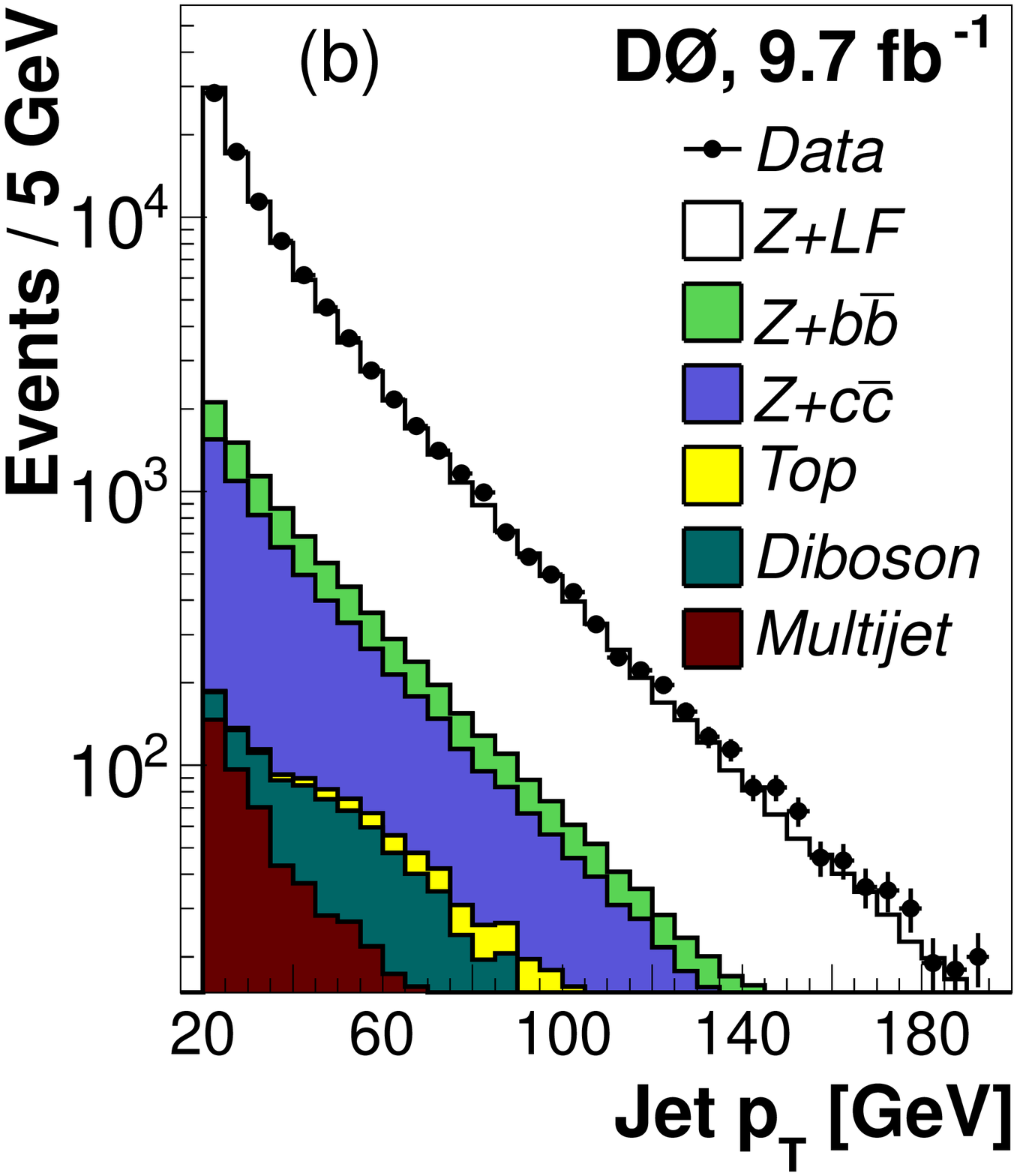}
\caption{\label{fig:jetpT_presel}(color online) The leading jet $p_{T}$ in the (a) $Z\rightarrow ee$
and (b) $Z\rightarrow \mu\mu$ channels for data and background in events with a $Z$ boson
candidate and at least one jet before $b$ tagging is applied.}
\end{figure}

The background contribution from multijet instrumental background events, 
in which jets are misidentified as leptons,
is evaluated from data. This is performed using a multijet-enriched sample of
events that pass all selection criteria except for some of the lepton quality
requirements. In the case of electrons, the multijet sample is obtained by
inverting the shower shape requirement and relaxing other electron
identification criteria, while for the muon channel, the multijet sample consists
of events with muon candidates that fail the isolation requirements. 
The normalization of the multijet background is
determined from a simultaneous fit to the dilepton invariant
mass distributions in different jet multiplicity bins. 
Figure~\ref{fig:jetpT_presel} shows the leading (in \pte) jet \pt distributions
compared to the expectations from various processes.
The dominant contribution comes from $Z+$light jet production.
The background fraction in the $ee$ channel is about 9.6\%,
and is dominated by multijet production. The muon channel 
has a higher purity with a background fraction of less than 1.3\%.

This analysis employs a two-step procedure to determine the $b$ quark 
content of jets in the selected events. First, a $b$-tagging algorithm
is applied to jets to select a sample of $Z+\text{jet}$ events
that is enriched in heavy flavor jets. After $b$ tagging, the relative light, charm, 
and $b$ quark content is extracted by fitting
templates built from a dedicated
discriminant that provides an optimized separation between the three components. 

Jets considered for $b$-tagging are subject to a preselection requirement,
called taggability, to decouple the intrinsic performance of the $b$ jet tagging
algorithm from effects related to track reconstruction efficiency. 
For this purpose, the jet is required to have at least two associated
tracks with \pt$>0.5~\GeVe$, the leading track must have
\pt$>1~\GeVe$, and each track must have at least one
SMT hit. This requirement has a typical efficiency of 90\% per $b$ jet.

The $b$-tagging algorithm is based on a 
multivariate analysis (MVA) technique~\cite{MVA}. This algorithm, MVA$_{bl}$, 
discriminates $b$-like jets from light-flavor-like jets
utilizing the relatively long lifetime of the $b$ hadrons when compared 
to their lighter counterparts~\cite{bid}. Events with at least one jet tagged 
by this algorithm are considered.

The MVA$_{bl}$ discriminant combines various properties of the jet and associated tracks
to create a continuous output that tends towards unity for $b$ jets and zero for
light jets. 
Inputs include the number of  secondary vertices and the 
charge track multiplicity, invariant mass of the secondary vertex ($M_{\text{SV}}$), decay length 
and impact parameter of secondary vertices, the charged 
tracks associated with them, and the Jet Lifetime Probability (JLIP), which
is the probability that tracks associated with the jet originate from the interaction vertex~\cite{bid}.
Events are retained for further analysis if they contain at least one jet with an MVA$_{bl}$
output greater than 0.1.
After these requirements, 8,042 $Z+\text{jet}$ events are
selected with at least one $b$-tagged jet, where only the 
highest \pt tagged jet
is examined in the analysis. The efficiency for tagging
$b$, $c$, and light jets are approximately 58.5\%, 19.8\%, and 2.41\%,
respectively. The resulting background 
contamination from diboson, multijet, and top production after $b$-tagging, 
for the electron and muon channels are 10.0\% and 3.6\%, respectively. 

\begin{figure}
\centering
 \includegraphics[width=0.50\columnwidth]{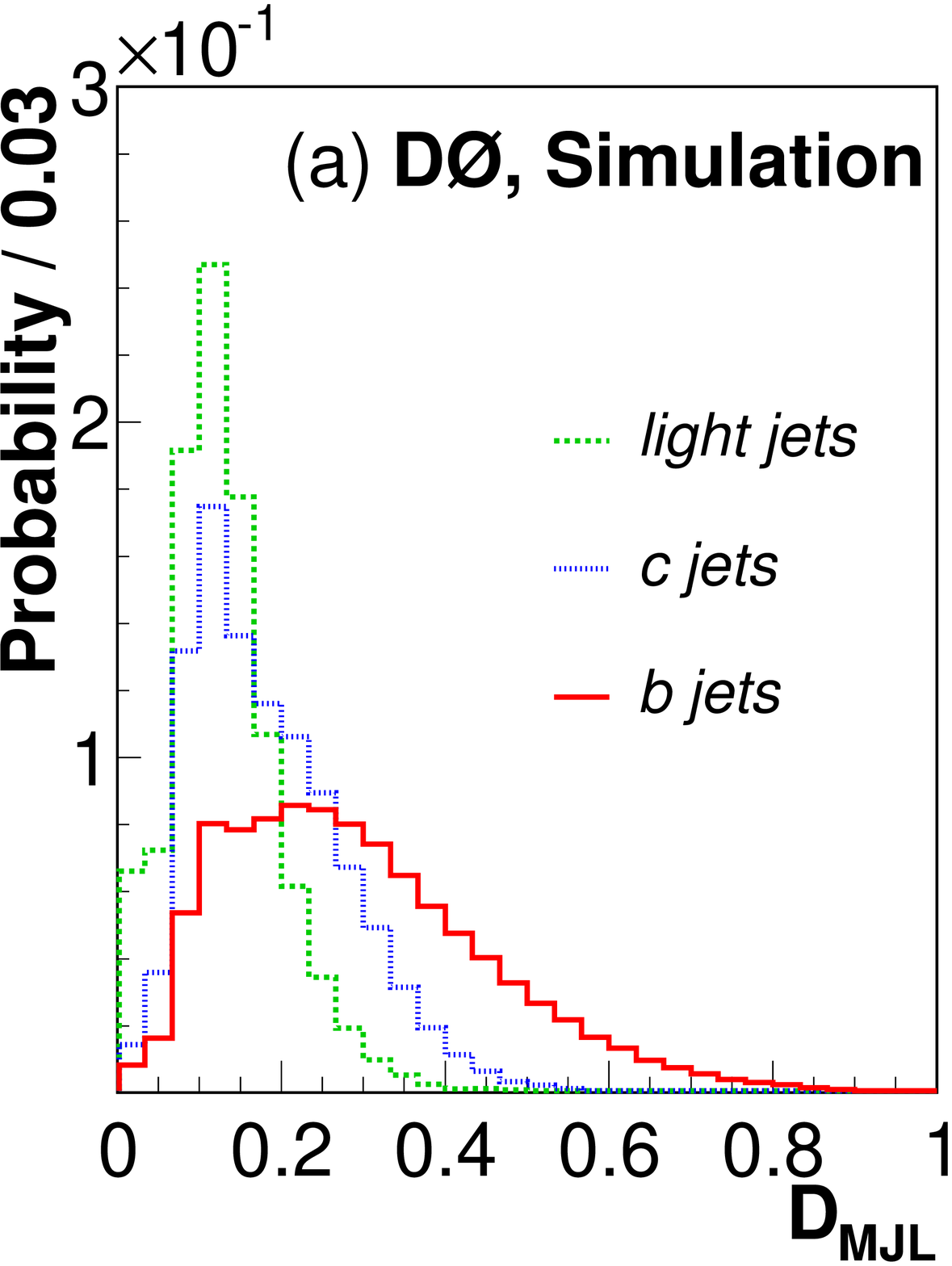}\includegraphics[width=0.50\columnwidth]{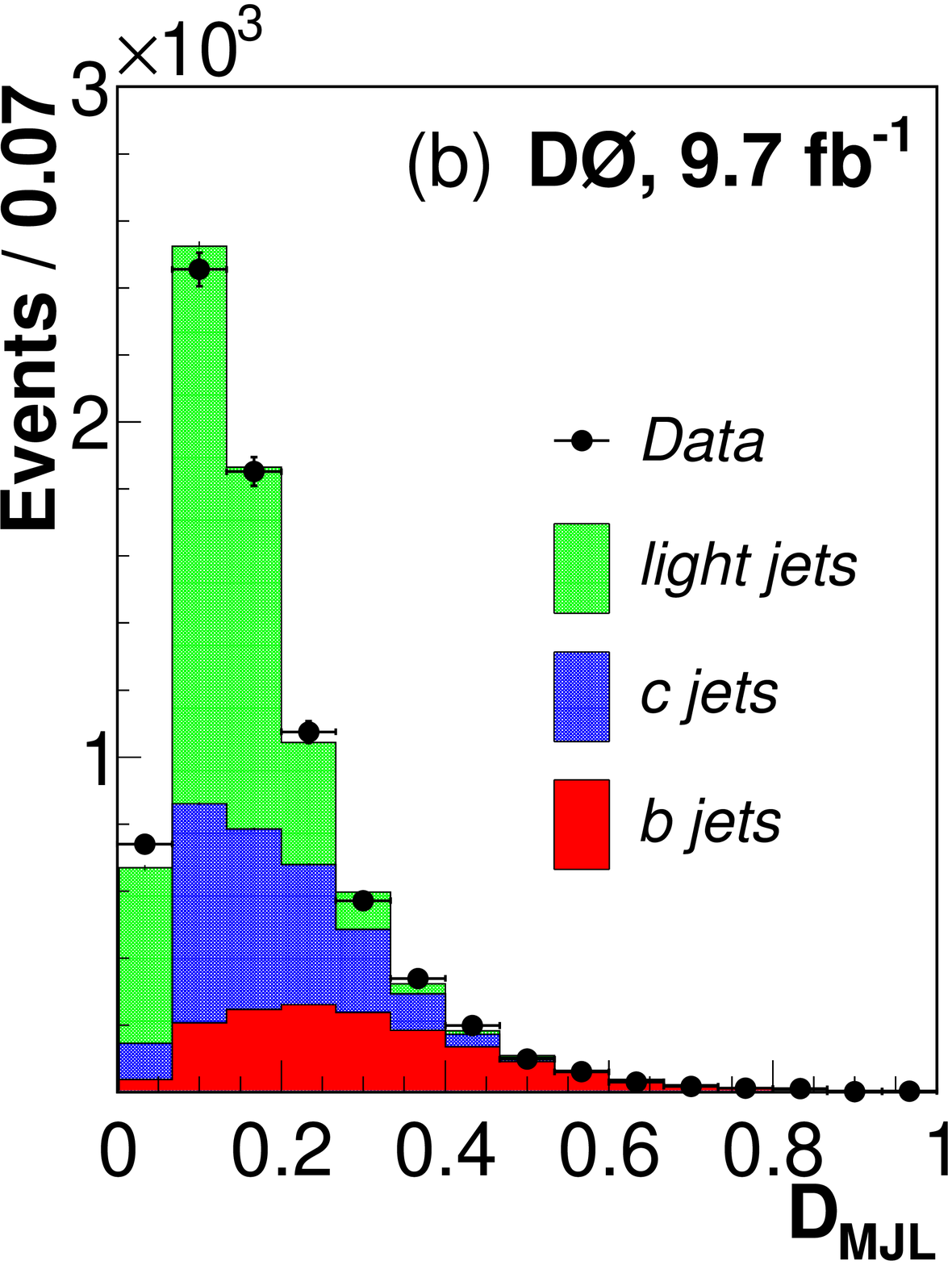}

\caption{\label{fig:channel_fit} (color online) 
(a) The probability densities of the \mjl discriminant for $b$, $c$,
and light jets passing the $b$ tagging requirements, normalized to unity. These templates are
taken from MC simulations. 
(b) The \mjl discriminant distribution of
events in the combined sample. The distributions of the $b$, $c$,
and light jets are normalized by the fractions found from the fit.}
\end{figure}

To determine the fraction of events with $b$, $c$ and light
jets, a dedicated discriminant, \mjle, is employed~\cite{Zb_PRD,Wb}. 
It is a combination of the two most discriminating MVA$_{bl}$ inputs, 
$M_{\text{SV}}$ and JLIP.
Figure~\ref{fig:channel_fit}(a) shows the
\mjl distributions (templates) obtained from simulations of all three considered
jet flavors that pass the $b$-tagging requirement.

To measure the fraction of events with different jet flavors
in the selected sample, we perform a binned maximum likelihood
fit to the \mjl distribution in data using the $b$, $c$, and light
flavor jet templates. Before the fit, all background 
contributions estimated after the MVA$_{bl}$ requirement, 
i.e., multijet, diboson and $t\bar{t}$ production,
are subtracted from the data leaving 3,576 and 3,921 $Z+\text{jet}$
events in the $ee$ and $\mu\mu$ channels, respectively.
Next, we measure the jet-flavor fractions in the dielectron and dimuon
samples separately, yielding the $b$ jet flavor fractions of 
$0.198\pm0.019~(\mbox{stat.})$ and $0.215\pm0.016~(\mbox{stat.})$, respectively. 
Since these measurements are in agreement within their statistical uncertainties,
we combine the two samples to increase the statistical power of
the fit for individual jet flavors. The measured fraction of $b$ jets in the combined sample 
is $0.207\pm0.011~(\mbox{stat.})$, the combined \mjl distribution of the 
$b$-tagged data and the fitted templates for the $b$, $c$, 
and light jets are shown in Fig.~\ref{fig:channel_fit}(b).

The fraction of $b$ jets measured in the heavy flavor enriched sample 
can now be combined with the corresponding acceptances for events 
to determine the ratio of cross sections using

\begin{eqnarray}
\frac{\sigma(Z+b~\text{jet})}{\sigma(Z+\text{jet})}=\frac{N\, f_{b}}{N_{\text{incl}}\,\epsilon_{tag}^{b}}\times\frac{\mathcal{A}_{\text{incl}}}{\mathcal{A}_{b}} ,
\label{eq:master}
\end{eqnarray}

\noindent where  $N_{\text{incl}}$ is the total number of $Z+\text{jet}$ events
before the tagging requirements, $N$ is the number of $Z+\text{jet}$
events used in the \mjl fit, $f_{b}$ is the extracted $b$ jet
fraction, and $\epsilon_{tag}^{b}$ is the overall selection efficiency of \mjl 
for $b$ jets which combines the efficiencies for taggability, MVA$_{bl}$
discriminant and \mjl selection. Both $N_{\text{incl}}$ and $N$
correspond to the number of events that remain after the contributions 
from non-$Z+\text{jet}$ processes have been subtracted from the data.
	
The detector acceptances for the inclusive jet sample ($\mathcal{A}_{\text{incl}}$)
and $b$ jets ($\mathcal{A}_{b}$) are determined from simulations in the kinematic region that
satisfies the $p_T$ and $\eta$ requirements for leptons and jets. 
The resulting ratio of the two acceptances is measured to be
$\mathcal{A}_{\text{incl}}/\mathcal{A}_{b}=1.118\pm0.002~(\mbox{stat.})$.
In this ratio, the effect of migration of events near the kinematic
threshold, or between neighboring kinematic bins, due to the finite detector
resolution is found to be negligibly small.

Using Eq.~(\ref{eq:master}), the result for the ratio of the $Z+b~\text{jet}$ cross section to the
inclusive $Z+\text{jet}$ cross section in the combined $\mu\mu$ and $ee$ channel
is $0.0196\pm0.0012~(\text{stat.})$.
In addition, the ratio \ratio~ 
of differential cross sections as a function of \ptj, \etaj, \ptz, 
and the azimuthal angle, \dphie, 
between the $Z$ boson and the highest \pt jet in the event, is measured. 
In these ratios the kinematics of the highest \pt $b$ tagged jet from the 
heavy flavor enriched sample is used in the numerator,
while the denominator of the ratio examines the kinematics of the highest \pt jet from 
the $Z+\text{jet}$ sample.
The data are split into five bins for each variable
such that the sample sizes allow for a stable
fit with the \mjl templates. The templates, in turn, are constructed
individually for every bin in the distribution of each examined variable. The selected bin sizes
along with the corresponding statistics of data events used in the fit are listed
in Table \ref{tab:Final}. In each case, all the quantities that
enter into Eq.~(\ref{eq:master}) are remeasured separately.
A summary of the differential cross section ratio measurements can
also be found in Table \ref{tab:Final}.

\begin{table}
\centering

\caption{\label{tab:Final} Results for the ratio \ratio~ in 
bins of \ptj, \ptz, \etaj, and $\Delta\varphi_{Z,\text{jet}}$. 
Bin centers, shown in parentheses, are chosen using the prescription found in Ref.~\cite{Terry}.}

\begin{tabular}{ccccc}
\hline 
\multirow{2}{*}{\ptj~{[}GeV{]}} & \multirow{2}{*}{$N$} & \multirow{2}{*}{$\frac{\sigma(Z+b~\text{jet})}{\sigma(Z+\text{jet})}$ } & Statistical  & Systematic\tabularnewline
 &  &  & Uncertainty & Uncertainty\tabularnewline
\hline
$20-30~(25)$  & 2920  & 0.0172 & 0.0014 & 0.0019\tabularnewline
$30-40~(35)$  & 1669 & 0.0210 & 0.0020 & 0.0016\tabularnewline
$40-55~(47)$  & 1358 & 0.0219 & 0.0022 & 0.0013\tabularnewline
$55-70~(62)$  & 616 & 0.0236 & 0.0035 & 0.0026\tabularnewline
$70-200~(102)$  & 915  & 0.0226 & 0.0042 & 0.0022\tabularnewline
\hline 
\ptz~{[}GeV{]} &  &  &  & \tabularnewline
\hline 
$0-20~(12)$  & 1066  & 0.0268 & 0.0028 & 0.0037\tabularnewline
$20-40~(32)$  & 2818  & 0.0119 & 0.0010 & 0.0010\tabularnewline
$40-60~(50)$  & 1925  & 0.0212 & 0.0019 & 0.0013\tabularnewline
$60-80~(68)$  & 887 & 0.0218 & 0.0031 & 0.0013\tabularnewline
$80-200~(100)$  & 789  & 0.0304 & 0.0050 & 0.0019\tabularnewline
\hline 
\etaj  &  &  &  & \tabularnewline
\hline 
$0-0.25~(0.13)$  & 1203  & 0.0139 & 0.0018 & 0.0010\tabularnewline
$0.25-0.5~(0.38)$  & 1207  & 0.0172 & 0.0017 & 0.0011\tabularnewline
$0.5-1.0~(0.75)$  & 2217  & 0.0213 & 0.0017 & 0.0017\tabularnewline
$1.0-1.5~(1.25)$  & 1695  & 0.0202 & 0.0020 & 0.0022\tabularnewline
$1.5-2.5~(2.00)$  & 1174  & 0.0161 & 0.0030 & 0.0023\tabularnewline
\hline 
$\Delta\varphi_{Z,\text{jet}}$ {[}rad{]} &  &  &  & \tabularnewline
\hline 
$0-2.5~(1.62)$  & 1612  & 0.0339 & 0.0037 & 0.0030\tabularnewline
$2.5-2.75~(2.63)$  & 957  & 0.0200 & 0.0027 & 0.0019\tabularnewline
$2.75-2.9~(2.83)$  & 1155  & 0.0210 & 0.0025 & 0.0017\tabularnewline
$2.9-3.05~(2.98)$  & 1937  & 0.0152 & 0.0015 & 0.0011\tabularnewline
$3.05-3.2~(3.13)$  & 1834  & 0.0129 & 0.0014 & 0.0008\tabularnewline
\hline 
\end{tabular}
\end{table}

Several systematic uncertainties cancel when the ratio 
\ratio~is measured. 
These include uncertainties
on the luminosity measurement, trigger, lepton, and 
jet reconstruction. The remaining uncertainties 
are estimated separately for the integrated result and in each bin
of the differential distributions.
For the integrated result, the largest systematic uncertainty of 5.3\% is due to  
the $b$ jet energy calibration; it comprises the uncertainties on the jet energy 
resolution and the jet energy scale. The 
next largest systematic uncertainty of 4.5\% comes from the shape of the \mjl templates used 
in the fit. The shape of the templates may be affected by the choice 
of the $b$ quark fragmentation function~\cite{topmass}, 
the background estimation, the difference in 
the shape of the light jet MC template and a template derived from a light jet enriched dijet data sample, 
the composition of the charm states used to determine the charm template shape~\cite{Zb_PRD}, 
and the uncertainty from the fit itself. These effects are evaluated by varying
the central values by the corresponding uncertainties, one at a time. 
The entire analysis chain has been checked for possible biases
using a MC closure test and no systematic effects has been found.
The other sources of uncertainty are due to
the $b$ jet identification efficiency (1.5\%) and the
choice of the MC event generator, {\sc alpgen} or {\sc pythia},
for the detector acceptance evaluations ($<0.1$\%). 
For the integrated ratio measurement, these uncertainties, 
when summed in quadrature, result in a
total systematic uncertainty of 7.1\%. 
The corresponding total systematic uncertainties for the ratios of differential cross sections are listed
in Table \ref{tab:Final}.
Finally, for the integrated \ratio~ratio we obtain a value of 
$0.0196\pm0.0012\left(\text{stat.}\right)\pm0.0013\left(\mbox{syst.}\right)$ which is in agreement with
the previous \dzero result of $0.0193 \pm 0.0027$~\cite{Zb_PRD}.  

\begin{figure*}
\centering

 \includegraphics[width =0.48\textwidth] {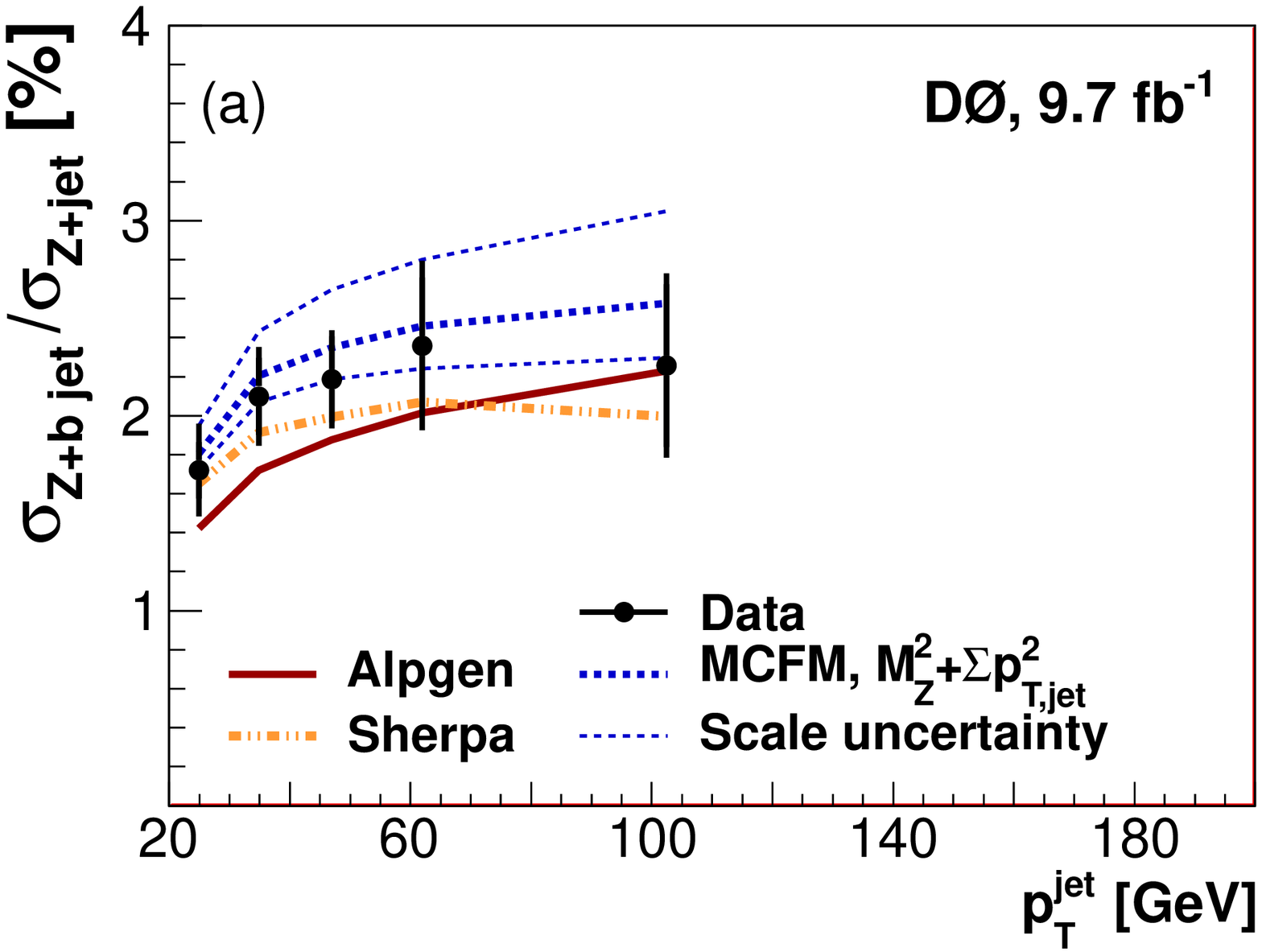} 
 \includegraphics[width =0.48\textwidth] {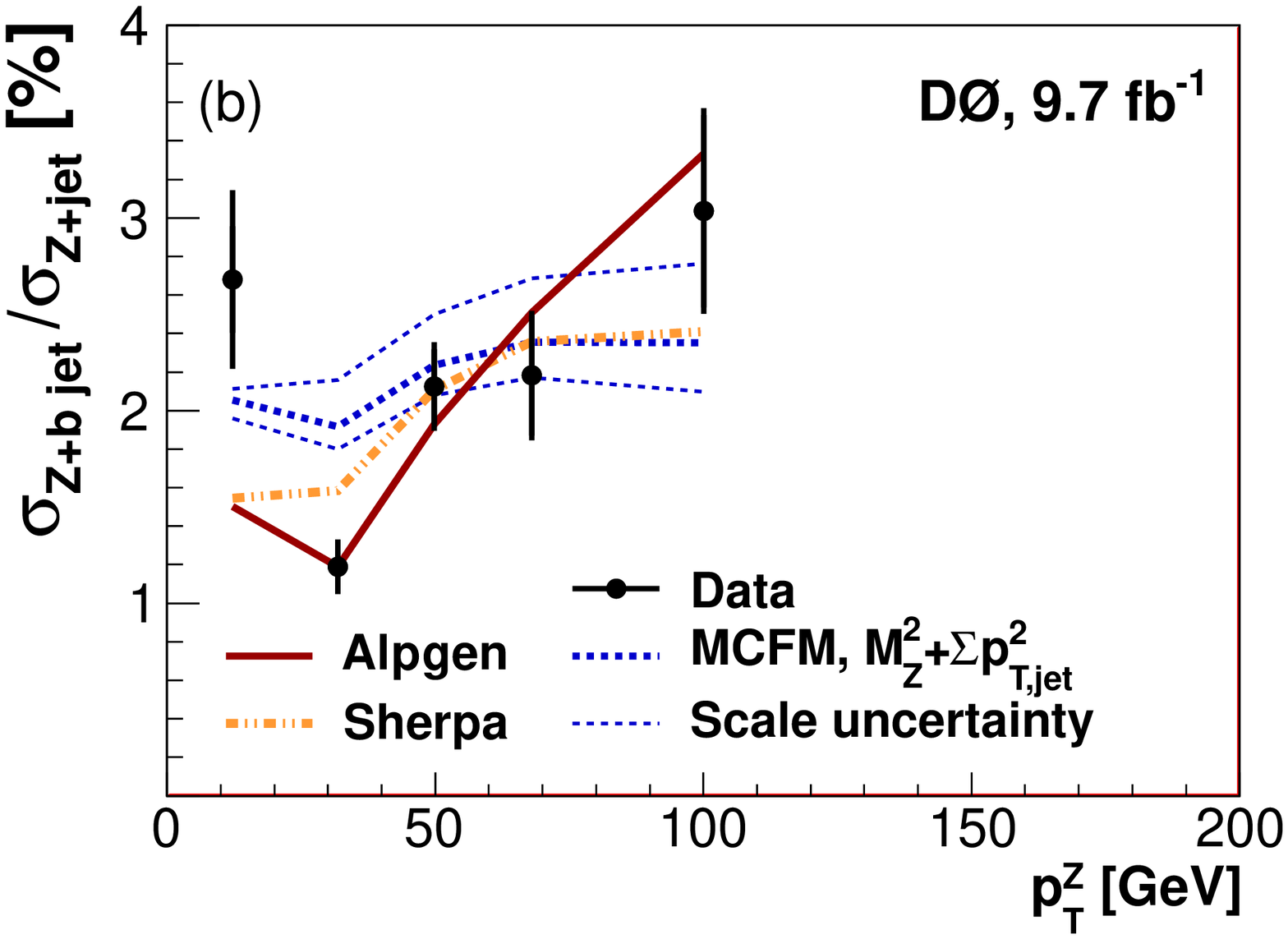}  \\
 \includegraphics[width =0.48\textwidth] {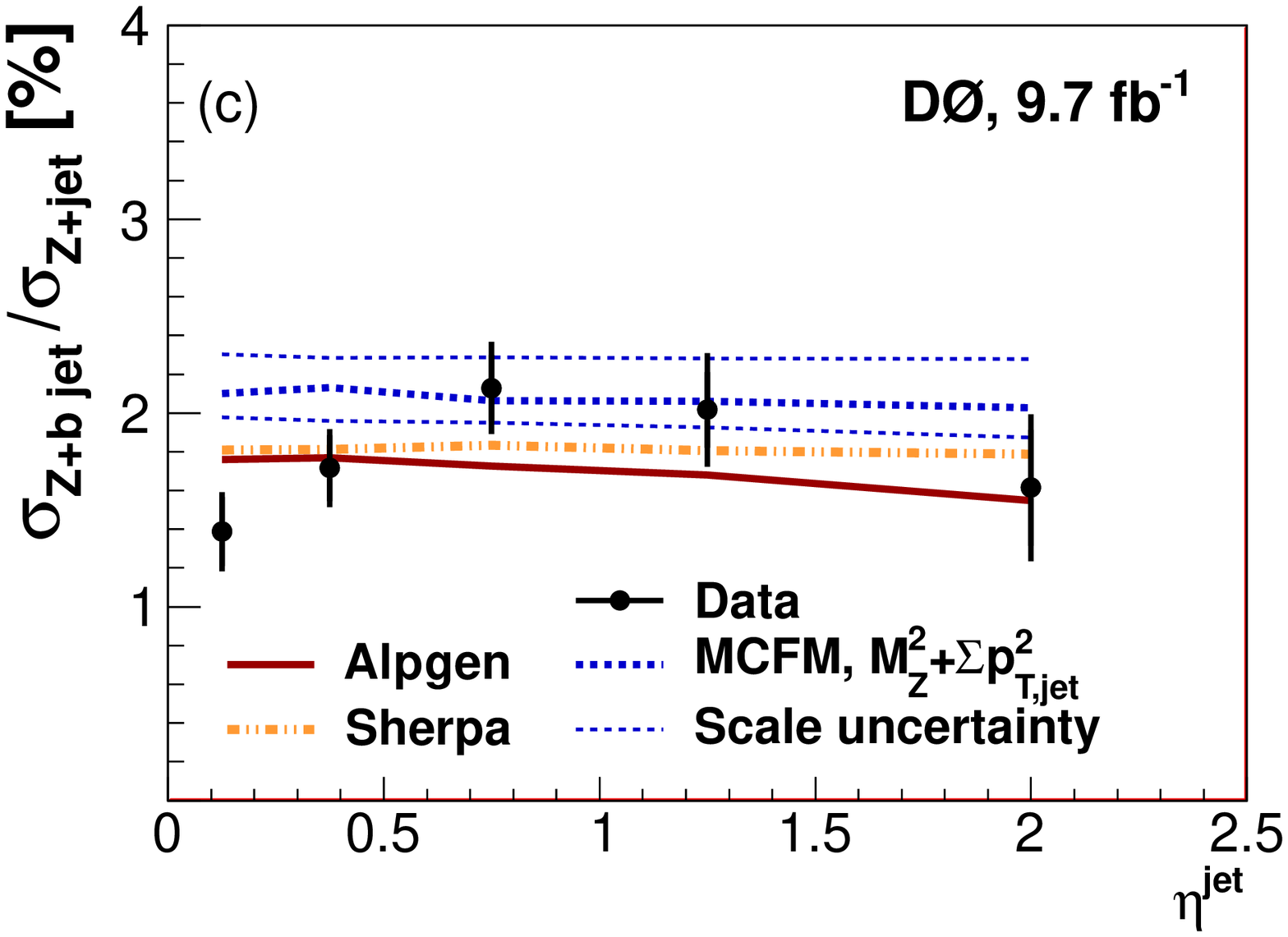} 
\includegraphics[width =0.48\textwidth] {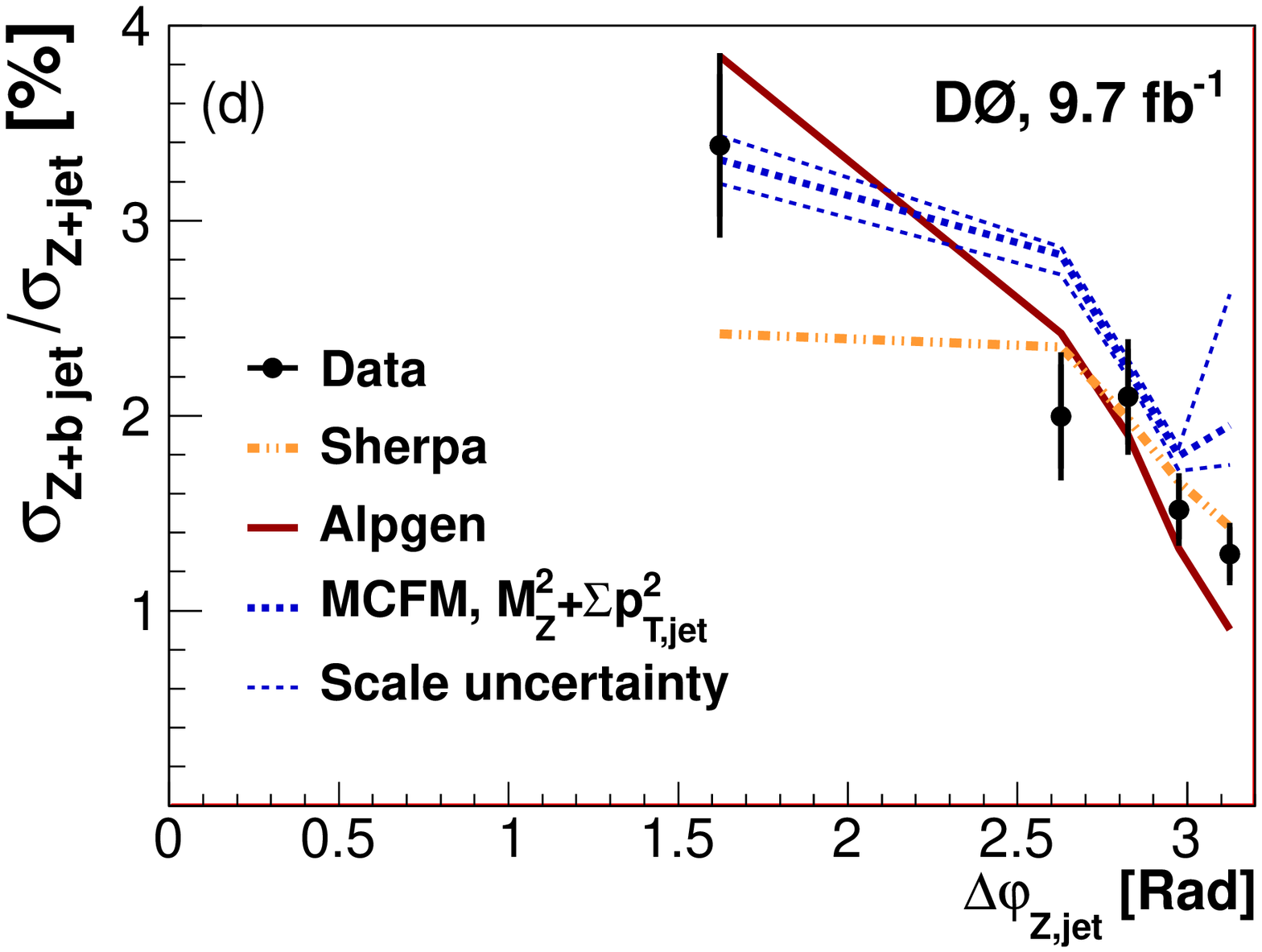} 

\caption{\label{fig:diffs} (color online) Ratios of the differential cross sections (a)  \ptj~(b) \ptz~(c) \etaj~and 
(d) \dphie. The uncertainties on the 
data include statistical and systematic uncertainties added in quadrature.  
The data are compared to the prediction from {\sc alpgen}, \sherpae, and the \mcfm NLO calculation, 
where the band represents the variation of the renormalization and factorization scales up and down by a factor of two. 
Bin centers are chosen using the prescription found in Ref.~\cite{Terry}.}

\end{figure*}

The measurements are compared to predictions from an NLO pQCD calculation
and two leading order MC event generators, \sherpa \cite{Sherpa} and \alpgene.
The NLO predictions are based on {\sc mcfm}~\cite{Campbell}, version 5.6,
with the MSTW2008 PDFs~\cite{mstw}
and the renormalization and factorization scales set at
$Q_{R}^{2}=Q_{F}^{2}=M_{Z}^{2}+\sum (\ptj)^{2}$. Here, $M_{Z}$ is 
the $Z$ boson mass and \ptj~is the transverse momentum of the jet(s). 
The measurement above is in agreement with the NLO pQCD prediction 
of $0.0206^{+0.0022}_{-0.0013}$~\cite{Campbell}, 
with corrections to account for non-perturbative effects estimated using {\sc alpgen+pythia}.
Uncertainties on the theoretical predictions are evaluated by
simultaneously changing the renormalization and factorization scales up
or down by a factor of two. 

Compared to an NLO calculation,
\sherpa uses the CKKW matching scheme between the leading-order matrix element
partons and the parton-shower jets following the prescription given in
Ref.~\cite{CKKW}. This effectively allows for
a consistent combination of the matrix element and
parton shower.

\alpgen also generates multi-parton final states using tree-level
matrix elements. When interfaced with {\sc pythia}, it
employs an MLM scheme~\cite{MLM} to match matrix
element partons with those after showering in {\sc pythia}, 
resulting in an improvement over leading-logarithmic
accuracy. 

The ratio of differential cross sections as a function of \ptj,
\ptz, \etaj, and \dphi are compared to predictions from {\sc mcfm}, \alpgene, and \sherpa 
in Fig.~\ref{fig:diffs}.
None of the predictions can fully describe all the examined
variables, except for the \ptj. Based on a $\chi^2$ test we find that 
the dependence on the \ptz~and \dphi correlation are
best described by \alpgen and \sherpae, respectively. Overall the integrated
result is best described by NLO predictions obtained with \mcfme. 

In summary, we have measured the ratio of integrated cross sections,
$\sigma(p\bar{p}\rightarrow Z+b~\text{jet})$/$\sigma(p\bar{p}\rightarrow Z+\text{jet})$,
as well as the ratio of the differential cross sections in bins of
\ptj, \ptz, \etaj, and $\Delta\varphi_{Z,\text{jet}}$,
for events with $Z\rightarrow\ell\ell (\ell = e,\mu)$ 
and at least one $b$ jet in the final state. Measurements are based on
the full data sample collected by the \dzero experiment in Run~II of the Tevatron,
corresponding to an integrated luminosity of 9.7 fb$^{-1}$ at
a center-of-mass energy of 1.96 TeV. For jets with
\ptj~$>20$~GeV and pseudorapidity $|$\etaj$|<2.5$, the measured integrated
ratio of $0.0196\pm0.0012\left(\mbox{stat.}\right)\pm0.0013\left(\mbox{syst.}\right)$
is in agreement with NLO pQCD predictions.
Results for the ratio of differential cross sections are also compared
to predictions from two Monte Carlo event generators. None of the 
predictions provide a consistent description of all the examined variables.

Supplementary material is available in~\cite{sup}.

%
We thank the authors of Refs.~\cite{Campbell, DawsonDoreen,Sherpa} for
valuable discussions, and the staffs at Fermilab, 
and collaborating institutions,
and acknowledge support from the
DOE and NSF (USA);
CEA and CNRS/IN2P3 (France);
MON, NRC KI and RFBR (Russia);
CNPq, FAPERJ, FAPESP and FUNDUNESP (Brazil);
DAE and DST (India);
Colciencias (Colombia);
CONACyT (Mexico);
NRF (Korea);
FOM (The Netherlands);
STFC and the Royal Society (United Kingdom);
MSMT and GACR (Czech Republic);
BMBF and DFG (Germany);
SFI (Ireland);
The Swedish Research Council (Sweden);
and
CAS and CNSF (China).


\begin{thebibliography}{10}
\bibitem{PV} The primary \ppbar interaction vertex is that found to be
the most likely collision point, among possibly several
collisions within a specific beam crossing, from which our
selected objects emanate. The algorithm for defining it
can be found in~\cite{bid}.
\bibitem{def} Pseudorapidity is defined as $\eta=-\ln[\tan(\theta/2)]$, with the polar
angle $\theta$ measured relative to the proton beam direction.
\end{thebibliography}
\end{document}